# Use Cases of Computational Reproducibility for Scientific Workflows at Exascale

Line Pouchard
*Center for Data Driven Discovery*
*Brookhaven National Laboratory*
Upton, NY
pouchard@bnl.gov

Sterling Baldwin
*Analytics and Informatics Management Systems*
*Lawrence Livermore National Laboratory*
Livermore, Ca
baldwin32@llnl.gov

Todd Elsethagen
*Data Sciences*
*Pacific Northwest National Laboratory*
Richland, Wa
todd.elsethagen@pnnl.gov

Carlos Fernando Gamboa
*Scientific Data and Computer Center*
*Brookhaven National Laboratory*
Upton, NY
cgamboa@bnl.gov

Shantenu Jha
*Center for Data Driven Discovery*
*Brookhaven National Laboratory*
Upton, NY
shanja@bnl.gov

Bibi Raju
*Data Sciences*
*Pacific Northwest National Laboratory*
Richland, Wa
bibi.raju@pnnl.gov

Eric Stephan
*Data Sciences*
*Pacific Northwest National Laboratory*
Richland, Wa
eric.stephan@pnnl.gov

Li Tang
*Center for Data Driven Discovery*
*Brookhaven National Laboratory*
Upton, NY
ltang@bnl.gov

Kerstin Kleese Van Dam
Computational Science Initiative
*Brookhaven National Laboratory*
Upton, NY 11973
kleese@bnl.gov

*Abstract*— We propose an approach for improved reproducibility that includes capturing and relating provenance characteristics and performance metrics, in a hybrid queriable system, the ProvEn server. The system capabilities are illustrated on two use cases: scientific reproducibility of results in the ACME climate simulations and performance reproducibility in molecular dynamics workflows on HPC computing platforms.

I. INTRODUCTION

The computational reproducibility of scientific experiments performed in diverse computing environments is both critical for the credibility of scientific results and difficult to achieve as numerous factors simultaneously impact the goals of reproducibility, how experiments are conducted, and the expected results. As workflows are more widely adopted in operational settings such as real time experimental data analysis, a secondary challenge is performance reproducibility, - can we reliably deliver the required results in a given tight time window? Both reproducibility challenges are strongly influenced by their computational conditions, which can refer to a litany of factors including: computational platform, system environment, reference data, application configuration, and third party libraries. The computing environments in which some of these experiments are conducted are adding to the reproducibility obstacles, in particular the DOE Leadership Class Facilities are built with multiple versions of unique libraries that evolve quickly, and furthermore they have a variety of hardware architectures that also continuously evolve. Compounding the problem is the fact that the multi-threading used in certain architectures can make program execution nondeterministic, and with performance variations resulting from the order in which memory is accessed on multi-core systems. To illustrate the challenges we will present two kinds of use cases addressing 1) the scientific reproducibility of computational results in the Accelerated Climate Modeling for Energy project [1] and 2) the lack of reproducibility of performance in computational workflows in large-scale environments. Our approach for improving reproducibility is based on provenance to simultaneously capture the experimental parameters and computing environment of these experiments. The goal is to extract the pertinent information, convert it into a unified format, and provide easy access. We will discuss the heterogeneity of relevant information, some tools to extract it, and the need for formalized computational workflows. The ProvEn server, a hybrid, queriable system for harvesting, storing and serving provenance will also be discussed [2].

## II. USE CASES

### A. Reproducibility of Scientific Results

Global climate models have a unique position in the physics simulation world in that they exist to both answer scientific questions as well as inform public policy. This coupled with the extremely large and complex nature of its subject matter means that reproducibility is absolutely critical to the success of the ACME project. There are two main driving factors behind the ACME project's needs. First, all data produced by the model must be independently verifiable or it loses all credibility. Second, the ACME development team is widely distributed across many research institutions and heterogeneous compute facilities. As the project develops and the code progresses, it's becoming increasingly apparent that every step to produce a piece of data absolutely must be accessible when viewing the model output. Due to the extremely computationally intensive nature of producing model output, careful consideration must be made when setting up a new run, which means previous successful run configurations greatly influence choices for future simulations. Historically much of this information has been captured manually and made available via shared documents, but as both the project team and code base grow this manual approach increasingly fails to robustly capture critical reproducibility data. When collecting run configuration data manually, member teams working on different submodules will collect different information, leaving us with irregular and incomplete data. Completely reproducing a run not only requires the input parameters such as surface temperature forcing, but also machine specifications such as compiler flags, environment variables, number of cores, as well as the hash of the code used.

### B. Reproducibilty of Performance

We investigated the reproducibility of two distinct workflow systems at progressively increasing scales. These workflows were drawn from biomolecular simulations and high-energy physics and were executed on DOE leadership class facilities. The former workflow is comprised of multiple concurrent executions of molecular dynamic (MD) simulations using the Gromacs MD engine (Fig.1); the high-energy physics workflows is constituted of multiple AthenaMP tasks (where each task runs on one node). Significant fluctuations in the total time to completion (TTC) between different runs of the same workflow were observed. The workflows were by design comprised of identical tasks, and thus the time to completion of each task should in principle have been identical. However, the individual and independent tasks of both workflows -- biomolecular and high-energy physics, had different execution times. Although the usage of the filesystem clearly contributes to the fluctuation in the second case, it is not the only contribution.

## III. APPROACH: INFORMATION EXTRACTION

For the ACME simulations, moving to an automated system will improve standardization of captured data, and allow for analysis of input parameters with fewer accidentally broken configurations and better utilization of machine allocations. The information captured by the ProvEn server includes git hash of the code used, the component set and resolution, machine and compiler details, file name lists and the job submission script. Scientists are encouraged to use a standardized "run_acme" script that captures this information, that is then ingested by the ProvEn provenance harvester.

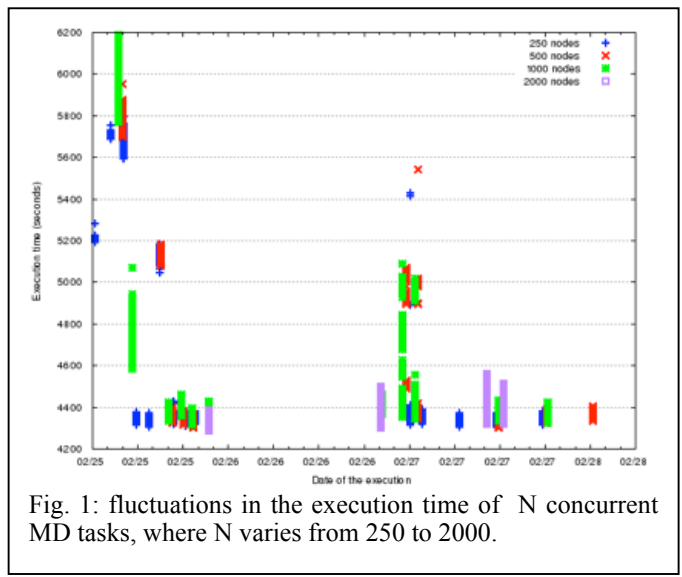

Fig. 1: fluctuations in the execution time of N concurrent MD tasks, where N varies from 250 to 2000.

For performance, we suggest that workflow systems - whether monolithic or comprised of building blocks [3] should specify information needed for reproducibility and enable its access. Tools capturing output from performance analysis tools and relating it to the execution environment, code branches, and libraries used have been designed [4]. Reproducibility necessitates a critical requirement to record the conditions under which experiments are performed. When simulations are fundamentally stochastic and have an additional contribution to the random variations in their times-to-completion, full "blanket" reproducibility will be difficult to achieve, and is best viewed as a multiple level capability characterized by different trade-offs and costs.